\documentclass{article}

\usepackage[preprint,nonatbib]{neurips_2023}

\usepackage{graphicx}
\usepackage[utf8]{inputenc} 
\usepackage[T1]{fontenc}    
\usepackage[symbol]{footmisc}
\usepackage{hyperref}       
\usepackage{url}            
\usepackage{booktabs}       
\usepackage{amsfonts}       
\usepackage{nicefrac}       
\usepackage{multirow}
\usepackage{tabularx}
\usepackage{caption}
\usepackage{amssymb}
\usepackage{amsmath}
\usepackage{bbold}                              
\usepackage{microtype}
\usepackage{siunitx}
\usepackage{etoolbox}                           
\newcommand{\ubold}{\fontseries{b}\selectfont}  
\robustify\ubold                                

\newcolumntype{C}{>{\centering\arraybackslash}X}
\newcolumntype{L}{>{\raggedright\arraybackslash}X}
\newcommand{\tablecaptionsep}{\vspace*{1pt}}

\usepackage{xcolor}

\title{Voice Conversion for Stuttered Speech, \\Instruments, Unseen Languages and\\ Textually Described Voices}

\author{%
  Matthew Baas \\
  \texttt{20786379@sun.ac.za} \\
   \And
   Herman Kamper \\
   \texttt{kamperh@sun.ac.za}\\
   \AND
  \textrm{\normalfont{MediaLab, Department of Electronic \& Electrical Engineering}}\\
  Stellenbosch University\\
  South Africa \\
}

\begin{document}

\maketitle

\begin{abstract} %
Voice conversion aims to convert source speech into a target voice using recordings of the target speaker as a reference.
Newer models are producing increasingly realistic output.
But what happens when models are fed with non-standard data, such as speech from a user with a speech impairment?
We investigate how a recent voice conversion model performs on  non-standard downstream voice conversion tasks. 
We use a simple but robust approach called k-nearest neighbors voice conversion (kNN-VC).
We look at four non-standard applications: stuttered voice conversion, cross-lingual voice conversion, musical instrument conversion, and text-to-voice conversion.
The latter involves converting to a target voice specified through a text description, e.g.\ ``a young man with a high-pitched voice''.
Compared to an established baseline, we find that kNN-VC retains high performance in stuttered and cross-lingual voice conversion.
Results are more mixed for the musical instrument and text-to-voice conversion tasks. E.g., kNN-VC works well on some instruments like drums but not on others.
Nevertheless, this shows that voice conversion models -- and kNN-VC in particular -- are increasingly applicable in a range of non-standard downstream tasks.
But there are still limitations when samples are very far from the training distribution.
Code, samples, trained models:
\href{https://rf5.github.io/sacair2023-knnvc-demo/}{https://rf5.github.io/sacair2023-knnvc-demo/}. %

\end{abstract}

\def\modelname{{kNN-VC}}

\section{Introduction} %

The conventional goal of voice conversion is to change an input utterance to seem as though it is spoken by a different speaker while retaining the same words~\cite{Mohammadi2017vc_def}.
Older models could only convert to and from speakers seen during training, while more recent models can also handle unseen speakers~\cite{vc_categories_liu2021any}. 
There are early indications that recent models can even deal with non-speech inputs, e.g.,\ using animal sounds as the target audio~\cite{dog_human_vc}.
To assess how far voice conversion models have come, this work explores how a recent model can be applied to several non-standard downstream tasks. 

Concretely, we apply the recent k-nearest neighbors voice conversion (kNN-VC) approach~\cite{baas2023knnvc} to four non-standard tasks.
kNN-VC is a simple method where each feature in the source speech is replaced with its nearest neighbors in the target data.
The approach is underpinned by a large self-supervised speech model that is used as the feature extractor~\cite{chen2022wavlm}.
Because no part of the model's design explicitly assumes that the source or target data is regular human speech, we believe that kNN-VC would be particularly well-suited to non-standard downstream voice conversion tasks.

We look at four tasks.
The first 
is stuttered voice conversion, where the target %
speech comes from a speaker with a stuttering speech impairment~\cite{stuttering_vc}.
Secondly, we use kNN-VC for cross-lingual voice conversion where the source and target speakers use different languages~\cite{vcc2020}. Both the input and output languages are also unseen during training.
Thirdly --\ going even further out-of-domain --\ we consider completely non-speech input: we apply kNN-VC to musical instrument conversion, where the source and target ``utterances'' come from different musical instruments.
Fourthly, we use kNN-VC to perform text-to-voice conversion where, instead of using a reference recording, the target voice is specified through a user-provided textual description, e.g.,\ ``an elderly woman with a velvety and resonant low voice''.
This last task is performed by training a small multi-layer perceptron
to map a vector representation of the target speaker description to a distribution over known speakers. %

On these four non-standard tasks, we compare kNN-VC to an existing established voice conversion system called FreeVC~\cite{freevc}.
We find that kNN-VC retains high performance in stuttered and cross-lingual conversion, outperforming the baseline in terms of similarity to the target speaker.
But quality is poorer in instrument and text-to-voice conversion, where kNN-VC doesn't consistently outperform the baselines;
despite the degraded performance, the conversions are still compelling considering how far out-of-domain music and text are for a model that has only seen speech during training.
Taken together, our experiments indicate that new voice conversion models -- and kNN-VC in particular -- are increasingly applicable to non-standard downstream tasks, with a single model able to perform a range of tasks that it wasn't trained for.
However, there are still limitations in generalization when the model is presented with data very far from the training distribution.
We invite the reader to listen to samples and use our code:
\href{https://rf5.github.io/sacair2023-knnvc-demo/}{https://rf5.github.io/sacair2023-knnvc-demo/}.

The paper is organized as follows. We first present the background of kNN-VC and the tasks we explore in Section~\ref{sec:2_knnvc}.
The four sections that follow then each investigates one of the tasks in detail. Finally, we conclude in Section~\ref{sec:conclusion}.

\section{Background}
\label{sec:2_knnvc} %

Voice conversion has a long history, but research has been accelerating in the last decade due to advances in generative machine learning~\cite{vc_overview_2021}.
The setup for most voice conversion methods is the same: the model takes in two inputs and produces one output.
The first input is the source speech which specifies the linguistic content to appear in the output, and the second is the reference speech which specifies the target speaker~\cite{Mohammadi2017vc_def}.
Most models %
attempt to learn %
to disentangle content from speaker identity, so that 
output can be produced with the same linguistic content as the source but sounding as though the words are spoken by the target speaker specified by the reference.

There are various methods to perform this disentanglement, ranging from annotated text for linguistic content alignment~\cite{yourtts_v162-casanova22a} to various forms of information bottlenecks~\cite{freevc, vqmivc_wang21n_interspeech, instancenormvc}.
We are interested in applying a well-performing voice conversion model to downstream applications, not in developing a new model ourselves.
We therefore use a particularly simple but robust method: kNN-VC.

\subsection{K-Nearest Neighbors Voice Conversion (kNN-VC)} \label{subsec:knnvc_model}

kNN-VC was inspired by ideas from concatenative speech synthesis, where an output waveform is produced by stitching together segments from different utterances~\cite{concatenative_vc_2007, cuta_concatenative_vc}.
But instead of using raw speech waveforms directly, kNN-VC relies on high-level speech features obtained from a large neural network.  
It specifically uses the WavLM encoder for feature extraction.
WavLM~\cite{chen2022wavlm} is a pretrained, self-supervised speech representation model that transforms input speech into a vector sequence, producing a single vector for every 20~ms of input audio.
Intuitively, the model is trained by masking parts of the input speech and then trying to predict what the vectors should be in the masked regions.
The result is an encoder that maps raw speech to features that are linearly predictive of various aspects of speech~\cite{chen2022wavlm}.
kNN-VC uses the activations from layer six of WavLM, which are linearly predictive of speaker identity~\cite{baas2023knnvc}.
It is these features that underpin the simple kNN-VC approach.

Voice conversion is performed in kNN-VC by mapping each WavLM vector from the source utterance to the mean of its $k$ nearest vectors from the reference utterance.
The resulting vector sequence is then converted back into a waveform using a vocoder~\cite{baas2023knnvc}, specifically a HiFi-GAN~\cite{hifi-gan} vocoder adapted to map WavLM features to audio.
If there are multiple reference utterances for the target speaker, the features %
from each are simply collated together into a %
pool of features known as the matching set.
While the WavLM encoder and HiFi-GAN vocoder have only seen speech during training, kNN-VC's design does not make any specific design assumptions restricting it to speech.
This, combined with the generality of the non-parametric kNN algorithm, makes kNN-VC a good choice for testing it on non-standard conversion tasks.

\subsection{Non-standard Voice Conversion Tasks} \label{subsec:tasks}

For each of the four tasks we explore, the conversion process %
can be formulated as a specific setting of a conversion model's source and reference data.
Concretely, the specification for the source, reference and output of a model for each task
is given in Table~\ref{tab:vc_setup}.
Here we give more background on each of the four tasks.

\textbf{Stuttered Voice Conversion.}
In our first task, we want to convert input speech so that it inherits the voice of a speaker having a stuttering impairment (as shown in Table~\ref{tab:vc_setup}).
Imagine that a student needs to create a presentation for their thesis, but the student suffers from a speech impairment.
If we could solve the stuttered speech conversion task, then someone else could read a prepared script which would then be converted to the student's voice.
This would result in an output resembling the student's voice but without stutters (since the source was read by a fluent speaker).
Prior efforts to correct stuttered speech %
have used energy correlations of speech segments to attempt to remove repetitions from stuttered speech~\cite{stuttering_correction2020}.
However, this approach attempts to directly edit stuttered speech into a non-stuttered form, 
while
we focus on stuttered voice conversion
which would enable
a stuttering speaker to map their voice onto an existing fluent utterance. 
Related efforts in~\cite{stuttering_vc} attempt to train a generative adversarial network to generate fluent speech for a given set of speakers  with a range of speech impairments.
But the approach struggles with stuttered speech because the model fails to appropriately change the prosody.
A voice conversion approach can address this since prosody is inherited from the source utterance (and not the stuttering reference).

\textbf{Cross-lingual Voice Conversion.}
Here
the source and reference utterances come from different, potentially unseen, languages (Table~\ref{tab:vc_setup}).
This is the most well-understood of the four voice conversion tasks,
with the Voice Conversion Competition in 2020~\cite{vcc2020} having a dedicated track for cross-lingual conversion.
Converting speech between languages unseen during training would improve access to speech technologies by speakers of low-resource languages.
But many existing methods
are limited to languages seen during training~\cite{vcc2020}.
Newer zero-shot conversion models like FreeVC~\cite{freevc} do not have this limitation but still struggle with completely unseen languages.
We are particularly interested in seeing whether voice conversion models can generalize to unseen languages.
In our experiments, we therefore use kNN-VC and FreeVC (both trained exclusively on English) to convert utterances between several non-English language pairs.

\begin{table}[!t]
    \renewcommand{\arraystretch}{1.2}
    \centering
    \caption{
        Voice conversion setup for the four non-standard downstream tasks.
    }
    \tablecaptionsep
    \small
    \label{tab:vc_setup}
    \begin{tabularx}{0.99\linewidth}{@{}
        L
        @{\hspace*{0.3cm}}
        L@{\hspace{0.3cm}}L@{\hspace{0.3cm}}L@{}}
    \toprule
    Task & Source & Target reference & Output  \\
    \midrule
    \textit{Stuttered conversion} & Non-stuttered \ speech &  Stuttered speech from desired speaker & Non-stuttered speech in desired voice \\
    \textit{Cross-lingual conversion} & Language A utterance & Language B utterance  & Language A utterance in voice B   \\
    \textit{Instrument conversion} & Music with instrument A & Music with instrument B & ``Content'' from A using instrument B \\
    \textit{Text-to-voice conversion} & Source utterance A & Text description of a speaker B & Utterance A in voice B   \\
    \bottomrule
    \end{tabularx}
\end{table}

\textbf{Musical Instrument Conversion.}
In this difficult task, we apply voice conversion to convert music played by one instrument to sound as though it is played by another. %
In this case, the source and reference are both non-speech pieces of music from a single instrument (Table~\ref{tab:vc_setup}).
Most prior work in this area --\ unsurprisingly --\ uses models trained on music.
The goal is typically to give a user control over the style of the synthesized audio.
E.g.,~\cite{music_conversion_vae} 
uses a variational autoencoder to model music style, while more recent
studies
like~\cite{barahona2023noisebandnet} use differentiable digital signal processing methods to 
allow a user to modify the output audio.
Our goal is not to achieve state-of-the-art music synthesis, but rather to assess how well a voice conversion approach (trained on speech audio and not music) is able to generalize to out-of-domain tasks.
So, for instrument conversion, we still use kNN-VC and again compare it to FreeVC -- both models having only \mbox{seen speech data.}

\textbf{Text-to-voice Conversion.}
The last %
task
is one where, instead of using a reference utterance to specify the target speaker, a textual description is provided, e.g.,\ ``an older man with a British accent and a pleasing, deep voice''.
This is known as text-to-voice conversion or prompt-to-voice conversion~\cite{prompt-to-voice_coqui}.
This task is much less explored than the others.
It was first introduced by Gölge and Davis~\cite{prompt-to-voice_coqui} in 2022, who trained
a specialized prompt-to-voice model.
Their work is proprietary and they also specifically considered a text-to-speech task where the target voice is provided as a text description rather than a reference utterance.
This %
motivates our introduction of a simple extension to kNN-VC to allow for textually described target voices in a voice conversion setting.

\section{Stuttering Experiments} %

For stuttered voice conversion, we perform two experiments: a comprehensive large-scale evaluation and a smaller-scale practical conversion of monologue recordings.
The former allows us to quantify kNN-VC's performance at stuttered conversion using the standard objective metrics of voice conversion, while the latter  
shows that it can be used in a real-world practical setting.

\subsection{Performance Metrics}\label{subsec:metrics}

When evaluating voice conversion performance, two categories of metrics are used: %
ones which measure how \textit{intelligible} the output is (i.e. how much of the lingustic content it retains form the source), and ones which measure the \textit{speaker similarity} to the target speaker~\cite{vcc2020}.

To measure intelligibility, we follow the same approach as in~\cite{softvc,baas2023knnvc}:
we assume we know the transcript of the source utterance, and then compare this to the transcript of the converted output.
The transcript of the output is obtained using an existing high-quality automatic speech recognition system.
We then compute a word/character error rate (W/CER) between the transcript of the converted output and the source: the lower the error rate, the better the conversion (i.e.,\ more intelligible).
The automatic speech recognition system we use is the pretrained Whisper \texttt{base} model using a decoding beam width of 5~\cite{whisper_radford2022robust}.

For speaker similarity, we compute an equal error rate (EER) as in~\cite{softvc}.
This metric is computed by calculating speaker similarity scores between pairs of real/real and real/generated utterances.
The real/real pairs are assigned a label of 1, and real/generated pairs a label of 0.
An equal error rate is then computed using the similarity scores and the labels.
The speaker similarity score between two utterances is defined the cosine distance between speaker embeddings computed using a pretrained speaker embedding model~\cite{xvector}.
The higher the EER, the less able we are to distinguish between the target speaker and converted output (i.e.,\ better speaker similarity), \mbox{up to a theoretical maximum of 50\%.}

\begin{table}[!b]
    \renewcommand{\arraystretch}{1.2}
    \centering
    \caption{
        Stuttered voice conversion performance, where the source is fluent speech and the target reference is from a speaker with a stuttering impairment.
    }
    \tablecaptionsep
    \label{tab:stuttering}
    \begin{tabularx}{0.7\linewidth}{@{}
        L
        @{\hspace{0.2cm}}
        S[table-format=4.2]@{\hspace{0.5cm}}
        S[table-format=3.2]@{\hspace{0.5cm}}
        S[table-format=3.1]@{}}
    \toprule
    Method & {\; WER $\downarrow$} & {\; CER $\downarrow$} & {EER (\%) $\uparrow$} \\
    \midrule
    \textit{dev-clean topline} & 4.63 & 1.60 & {---} \\
    FreeVC & \ubold 6.59 & \ubold 2.57 & 25.6 \\
    kNN-VC & 14.61 & 7.16 & \ubold 46.7 \\
    \bottomrule
    \end{tabularx}
\end{table}

\subsection{Stuttering Events in Podcasts}

We evaluate kNN-VC and FreeVC on a combination the LibriSpeech's \texttt{dev-clean} dataset~\cite{panayotov2015librispeech} and the Stuttering Events in Podcasts dataset~\cite{lea_stutterdataset_sep28k}.
LibriSpeech consists of hundreds of hours of read books in English by speakers without any speech 
impairment,
while the Stuttering Events dataset consists
of data from podcasts where stuttering events have been annotated.
Both datasets 
contain speech from many speakers,
none of which are seen by kNN-VC or FreeVC.
Using the annotations in the Stuttering Events dataset, we trim each utterance to only those segments where there is a sound or word repetition,
i.e.,\ we filter out all non-stuttered speech and only use stuttered instances as reference audio.

As explained in Section~\ref{subsec:tasks}, we want the generated speech to match the identity of a reference speaker that stutters, but the content should match a fluent source utterance.
So, for our evaluation, we convert each of the 200 utterances from LibriSpeech's \texttt{dev-clean} subset from the kNN-VC paper~\cite{baas2023knnvc} to ten random %
speakers from the Stuttering Events dataset.
Since each reference speaker has several short clips with stuttered speech, 
we sample up to 30 stuttered reference clips for each target speaker.
These clips then serve as the matching set for conditioning kNN-VC when it is presented with a source utterance from LibriSpeech (see Section~\ref{subsec:knnvc_model}).
For the FreeVC~\cite{freevc} baseline, we
use the same input utterances
but compute the mean speaker embedding of the stuttering clips
to serve as the target.
This is done because FreeVC's target conditioning mechanism doesn't use a matching set; instead, it uses a trained speaker encoder which maps a reference utterance to a single speaker embedding~\cite{freevc}.

The result of this evaluation
are presented in Table~\ref{tab:stuttering}.
The topline is an approach where we perform no conversion and simply evaluate the baseline transcription error rates of the Whisper model on the ground truth source utterances.
We make two key observations from the results.
First, voice conversion can be used effectively for this task, with high EER (good speaker similarity) and low W/CER scores (good intelligibility).
Second, 
there is a tradeoff between kNN-VC and FreeVC:
FreeVC retains more of the desired linguistic content, with W/CERs that are less than half of the rates achieved by kNN-VC.
Conversely, %
kNN-VC is substantially
better in matching the target speaker, with a much higher EER.
We suspect that the better speaker similarity of kNN-VC is due to the matching set mechanism that is more robust to non-standard speech compared to the speaker embedding approach of FreeVC.
However, the matching approach of kNN-VC might also be what causes poorer intelligibility because the matching set contains several non-fluent phones and biphones.

\subsection{Practical Evaluation}\label{subsec:practical_stutter} %

Having objectively evaluated kNN-VC's performance, we now apply it in two practical scenarios in a qualitative evaluation.
First, we apply it to the example already mentioned in Section~\ref{subsec:tasks}, where we want to generate the audio of a thesis presentation for a student that suffers from a stuttering impairment.
The source to kNN-VC is provided by a fluent speaker reading a script provided by the student.
In this case, the reference data was provided by the student, but stuttering was removed by hand.
Second, we apply kNN-VC using stuttered speech from an actor as reference, specifically the actor Colin Firth acting as King George VI in the movie ``The King's Speech''.
In both cases, to further improve output quality, we also apply the \texttt{voicefixer} upsampling model to remove some %
conversion artifacts~\cite{voicefixer}.
We encourage the reader to listen to conversion samples at
\href{https://rf5.github.io/sacair2023-knnvc-demo/}{https://rf5.github.io/sacair2023-knnvc-demo/}.

\section{Cross-lingual Voice Conversion} %

In cross-lingual voice conversion, we want to convert to a target speaker talking a different language from the source.
The languages and identities of the source and target speakers are unseen during training.
We again apply kNN-VC as our main model and use FreeVC as a baseline.
Because the inputs and outputs in this task are still human speech, we  use the same intelligibility and speaker similarity metrics as in Section~\ref{subsec:metrics}. %

\subsection{Experimental Setup}

We conduct a large-scale evaluation on the Multilingual LibriSpeech dataset~\cite{mls_multilinguallibri_Pratap2020L}.
The dataset consists of thousands of hours of audiobook recordings across eight European languages.
For evaluation, we sample 16 utterances from three random speakers from the test subset of each language, yielding 384 evaluation utterances.
Evaluation involves converting each of these utterances to every other speaker in the evaluation set, giving a total of just under 9000 unique source-target evaluation pairs covering all possible language combinations.

For measuring intelligibility as described in Section~\ref{subsec:metrics}, we can still use Whisper \texttt{base} to transcribe the original and converted utterances because Whisper is a multilingual speech recognition model that can be presented with input in any of the European languages considered here~\cite{whisper_radford2022robust}.
Similarly, speaker similarity is evaluated using the same speaker embedding model, since these models are trained to ignore linguistic content and only extract speaker information.
We can therefore calculate an EER exactly as in Section~\ref{subsec:metrics}.

\begin{table}[!t]
    \renewcommand{\arraystretch}{1.2}
    \centering
    \caption{
        Cross-lingual voice conversion performance on the Multilingual Librispeech test dataset.
    }
    \tablecaptionsep
    \label{tab:cross-lingual}
    \begin{tabularx}{0.7\linewidth}{@{}
        L
        @{\hspace{0.2cm}}
        S[table-format=3.1]@{\hspace{0.5cm}}
        S[table-format=3.1]@{\hspace{0.5cm}}
        S[table-format=3.1]@{}}
    \toprule
    Method & {\; WER $\downarrow$} & {\; CER $\downarrow$} & {EER (\%) $\uparrow$} \\ %
    \midrule
    \textit{test set topline} & 21.5 & 7.1 & {---} \\ %
    FreeVC & 34.1 & 13.5 & 7.7 \\ %
    kNN-VC & \ubold 33.9 & \ubold 13.1 & \ubold 25.0 \\ %
    \bottomrule
    \end{tabularx}
\end{table}

\subsection{Results}

The cross-lingual voice conversion results on the Multilingual LibriSpeech dataset are presented in Table~\ref{tab:cross-lingual}.
Unlike the stuttered conversion experiments, we see that kNN-VC is consistently more intelligible and more similar to the target speaker than FreeVC.

How do cross-lingual conversion compare to converting between speakers in English (the language on which both FreeVC and kNN-VC are trained)?
If we compare the results in Table~\ref{tab:cross-lingual} to the monolingual scores reported in~\cite{freevc,baas2023knnvc}, we see that both the intelligibility and speaker similarity is lower when performing cross-lingual voice conversion compared to monolingual conversion.
This degradation is to be expected, given that these voice conversion models might not have seen all the phones necessary for converting between unseen language pairs.
E.g., 
when converting Polish to Spanish with kNN-VC, the Spanish matching set might not have all the phones necessary to represent the source Polish utterance.

Nevertheless, for kNN-VC in particular, the cross-lingual performance is still compelling, with the CER and EER scores in Table~\ref{tab:cross-lingual} coming close to FreeVC's performance on English~\cite{freevc}.
This shows that voice conversion models trained only on English  can be applied succesfully to unseen languages.
We encourage the reader to listen to a %
selection of conversion samples on our demo website.

\section{Musical Instrument Conversion} %

Moving on to harder tasks, we now consider
a conversion problem involving  non-speech audio.
Specifically, we want to see whether we can use kNN-VC to convert a piece of music played with one instrument to sound as though it is played with another -- without fine-tuning or adapting the model.
To this end, we use a dataset consisting of several music pieces, each played on a single instrument.
We then attempt to convert a song from one instrument to another by using all the audio from the target instrument as our matching set in kNN-VC.
This effectively treats the recordings from the target instrument as ``utterances'' from a target ``speaker''.

\subsection{Experimental Setup}

We use the Kaggle Musical Instrument's Sound Dataset~\cite{musical_instruments_dataset}.
The dataset consists of short music pieces played with one of four instruments:
drums, violin, piano, or guitar. %
For our music conversion evaluation set, we only consider recordings that are between 10 and 90~seconds long.
Evaluation involves converting each recording in the test set to every other instrument, resulting in 153 synthesized output audio recordings that are evaluated.
For kNN-VC, all the recordings from the target instrument are used as the matching set.
For FreeVC, a mean ``speaker'' embedding is obtained from the target instrument audio.

The intelligibility and speaker similarity metrics from Section~\ref{subsec:metrics} are not applicable to instrument conversion.
We therefore use a metric from the music synthesis domain: Fréchet audio distance (FAD)~\cite{Kilgour2019_fad}.
This metric uses a music classification model {(trained to classify the tags of scraped YouTube videos) to compare two recordings, giving a single number which roughly indicates how similar the music in the recordings is~\cite{Kilgour2019_fad}.
This is achieved by computing an ``Inception score'' between the intermediary features produced by the classifier for two sets of audio inputs.
In our case we measure the FAD between the converted outputs and the ground truth recordings from the target instrument.
E.g., to measure FAD for drums, we compare all the outputs where drums are the target instrument to the real recordings of drums.
We use an open-source implementation of FAD\footnote{\href{https://github.com/gudgud96/frechet-audio-distance}{https://github.com/gudgud96/frechet-audio-distance}}.

\subsection{Results}

\begin{table}[!b]
    \renewcommand{\arraystretch}{1.2}
    \centering
    \caption{
        Musical instrument conversion performance on the Musical Instrument's Sound Dataset, measured in FAD~\cite{Kilgour2019_fad}.
    }
    \tablecaptionsep
    \label{tab:music-instrument}
    \begin{tabularx}{0.7\linewidth}{@{}
        L
        @{\hspace{0.2cm}}
        S[table-format=4.2]@{\hspace{0.5cm}}
        S[table-format=4.2]@{\hspace{0.5cm}}
        S[table-format=4.2]@{\hspace{0.3cm}}
        S[table-format=4.2]@{}}
    \toprule
    &  \multicolumn{4}{c}{{Target instrument FAD $\downarrow$}}  \\
    \cmidrule(l){2-5}\vspace{-0.3cm}
    Method & {\; Drums} & {\; Violin} & {\; Piano} & {\; Guitar} \\
    \midrule
    FreeVC & 20.11 & \ubold 21.07 & \ubold 20.03 & 23.22 \\
    kNN-VC & \ubold 9.81 & 22.44 & 23.95 & \ubold 18.01 \\
    \bottomrule
    \end{tabularx}
\end{table}

The musical conversion performance for each instrument is presented in Table~\ref{tab:music-instrument}.
FAD values greater than $10$ are considered to be very poor~\cite{Kilgour2019_fad}.
So we can see that FreeVC performs very poorly, regardless of the instrument.
kNN-VC also performs poorly for most target instruments, except when converting to drums where the best overall FAD is achieved.
While a FAD of 9.8 is still poor compared to typical scores in music enhancement~\cite{Kilgour2019_fad}, qualitatively kNN-VC's conversion to drums sounds much closer to the target than any of the other conversions.

We are still unsure why kNN-VC performs so much better when converting to drums.
We speculate that this might be because drum audio has lower frequency content that might be in a similar range to that of human speech --\ at least compared to the high-frequency content from guitars, pianos, and violins.
The encoder and vocoder of kNN-VC have only seen speech, and it is therefore likely that the WavLM features would be better at capturing instrument information with a similar frequency range to speech.

Overall, the results indicate that the kNN-VC and FreeVC cannot perform instrument conversion, except in specific instances.
This therefore remains an open question for future work;
one idea would be to incorporate some of the techniques used in music enhancement and synthesis~\cite{barahona2023noisebandnet, music_conversion_vae}.

\section{Text-to-voice Conversion}\label{sec:text-to-voice} %

The last task we look at is text-to-voice conversion, where the target speaker is specified using a textual description instead of a reference utterance.
Unlike the previous tasks, kNN-VC cannot be directly used with text inputs.
So we propose a small extension.

\begin{figure}[!b]
    \centering
    \includegraphics[width=0.9\linewidth]{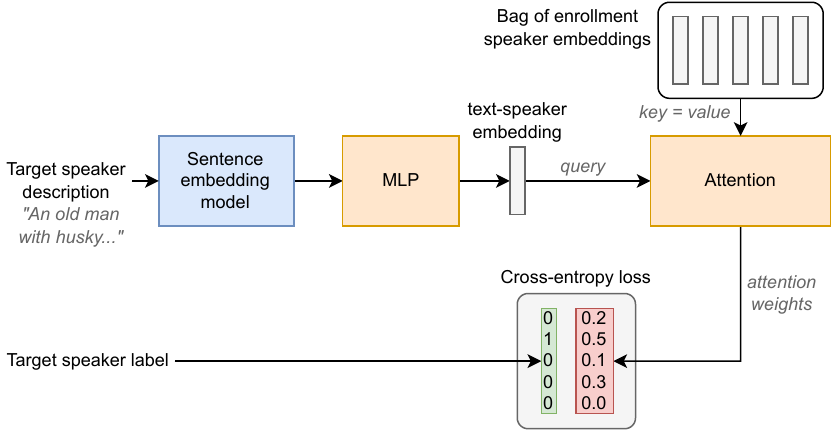}
    \caption{
    For text-to-voice conversion, we use an approach where a text description of a voice is converted to a distribution over enrollment speakers.
    Only the multi-layer perceptron (MLP) and attention layer are trained; the sentence embedding model and speaker embeddings are fixed.
    The parameters are optimized to predict the target speaker label from the provided textual description.
    }
    \label{fig:text-to-voice_diagram}
\end{figure}

\subsection{Extending kNN-VC for Textual Voice Descriptions}

We extend kNN-VC to work with text descriptions using an insight from Cheng~\cite{cheng_knnvc_interpolation}: 
during synthesis with kNN-VC,
we can construct combinations of voices by linearly interpolating the converted WavLM features between those from different target speakers, before vocoding.
We therefore rephrase the text-to-voice conversion task as finding a probability distribution over a set of enrollment speakers that best matches the provided text description.
If we knew this distribution, we could use the probabilities as weighting coefficients for interpolating the converted output to best match the target speaker description. 

Fig.~\ref{fig:text-to-voice_diagram} illustrates our approach for predicting the distribution over speakers from a target speaker description.
We construct a small text-to-voice matching network that maps a text description to  a distribution over a set of enrollment speakers.
The network consists of three components:

\begin{enumerate}
    \item A sentence embedding model~\cite{reimers-2019-sentence-bert}, which converts a text description into a single vector capturing the target speaker identity.
    We use the pretrained model from~\cite{reimers-2019-sentence-bert} directly.
    \item A bag of speaker embeddings computed using a pretrained speaker embedding model~\cite{GE2E};
    a single speaker embedding is obtained for each enrollment speaker by averaging the embeddings from that speaker's utterances.
    We use an open-source pretrained speaker embedding model~\footnote{\href{https://github.com/RF5/simple-speaker-embedding}{https://github.com/RF5/simple-speaker-embedding}}.
    \item 
    A multi-layer perceptron (MLP) with three layers and an attention~\cite{vaswani2017attention} head. The MLP maps the sentence embeddings~\cite{reimers-2019-sentence-bert} to a ``text-speaker embedding'', which is used as the query over the bag of enrollment embeddings (acting as keys and values in the attention operation).
\end{enumerate}

The output of this small matching network is the attention weights over the keys/values, defining a distribution over the enrollment speakers.
During training, the model is optimized using a cross-entropy loss  between the one-hot target speaker label and the predicted distribution over speaker embeddings.
During inference, the distribution is used as the weighting for interpolating the converted WavLM output features of kNN-VC before vocoding.

\vspace{-2mm}
\subsection{Experimental setup}

We use the LibriSpeech~\cite{panayotov2015librispeech}  \texttt{train-clean-100} subset for this experiment.
We manually labelled 100 speakers with a textual description of the speaker's voice, e.g.,\ ``a young woman with a squeaky and animated voice, speaking in a rapid and highly expressive manner''.
We split this set into 90 training and 10 test speakers.
The MLP and attention layer from Fig.~\ref{fig:text-to-voice_diagram} is then trained for 300 steps using AdamW~\cite{adamw} with a learning rate of $5\cdot 10^{-5}$ and weight decay of $3\cdot 10^{-3}$.
The MLP consists of three layers with LeakyReLU activations and layer normalization~\cite{ba2016layernorm} between each layer.
Since the sentence and speaker embeddings respectively have 384 and 256 dimensions, the output dimensions from the three MLP layers are 1024, 768 and 256.
The attention is a standard scaled dot-product multi-head attention with 16 heads~\cite{vaswani2017attention}.

The evaluation task involves converting the same set of \texttt{test-clean} utterances as in~\cite{baas2023knnvc} to each of the ten test speaker descriptions.
First the network in Fig.~\ref{fig:text-to-voice_diagram} maps the description of each test target speaker to a distribution over the 90 enrollment  speakers. 
Then we perform kNN-VC conversion to each enrollment speaker as target.
Finally, we interpolate the output WavLM features in the ratio provided by the attention weights of the model in Fig.~\ref{fig:text-to-voice_diagram} before vocoding to achieve the final converted output.
Using the set of converted outputs, speech inputs, and original reference utterances, we assess performance with the same metrics as in Section~\ref{subsec:metrics}.

\subsection{Results}

The text-to-voice conversion results are presented in Table~\ref{tab:text-to-voice}.
We see that intelligibility is not harmed when using textually described target voices --\ 
it actually improves slightly compared to the 1-minute topline.
This makes sense since the target is based on a combination of the training speakers, for which we have much more data to use as matching data compared  to the topline.
However, this comes at a cost: the similarity to the target speaker is much worse compared to the topline (lower EER).
While not certain, we hypothesize this is largely due to the highly subjective nature of mapping utterances from a speaker to a textual description of that speaker's voice.
E.g.,\ what one person might label as a ``squeaky and animated voice'' might differ greatly from how another person describes the same voice.
This is compounded by the sensitivity of the speaker embedding model used to evaluate EER, since it is trained to be sensitive enough to discern even similar speakers of the same gender and age.
Overall, the results indicate that we can perform text-to-voice conversion to some degree, but there is still more work required to accurately map to the desired target speaker.

\begin{table}[t]
    \vspace{-5pt}
    \renewcommand{\arraystretch}{1.2}
    \centering
    \caption{
        Text-to-voice conversion performance on LibriSpeech,
        where unseen source speakers are converted to target speakers specified with a textual description.
        Topline performance when 1 minute of speech audio is used to specify the target speaker (instead of text) is provided as comparison. 
    }
    \tablecaptionsep
    \label{tab:text-to-voice}
    \begin{tabularx}{0.9\linewidth}{@{}
        L
        @{\hspace{0.2cm}}
        S[table-format=3.1]@{\hspace{0.5cm}}
        S[table-format=3.1]@{\hspace{0.5cm}}
        S[table-format=3.1]@{}}
    \toprule
    Method & {\; WER $\downarrow$} & {\; CER $\downarrow$} & {EER (\%) $\uparrow$} \\ %
    \midrule
    \textit{1~min speech kNN-VC topline} & 7.73 & 3.09 & \ubold 35.05 \\
    kNN-VC three-layer MLP & \ubold 6.32 & \ubold 2.51 &  3.15 \\ %
    \bottomrule
    \end{tabularx}
\end{table}

\vspace{-0.2cm}
\section{Conclusion}~\label{sec:conclusion} %
\vspace{-0.5cm}

We explored the generalization capabilities of voice conversion models by applying a specific model (kNN-VC) to several non-standard conversion tasks.
In stuttered voice conversion and cross-lingual voice conversion, kNN-VC retained high performance,
achieving convincing results compared to a baseline.
For music conversion, however, the results were more mixed, with kNN-VC being largely unable 
to produce high-quality conversions except when converting to drums.
The results for textually described voice conversion were also mixed: the output was still intelligible but was a poor match to the target speaker compared to a standard system using audio rather than text to specify the target speaker.
Taken together, we showed that an existing voice conversion system can be applied to a range of non-standard downstream voice conversion tasks without major changes to the model.
Future work will consider how voice conversion approaches can be improved to explicitly deal with inputs that are very far from the data they have been exposed to during training.

\vspace{-0.2cm}
\subsubsection{Acknowledgements.}

We would like to thank Mikkel du Plessis for feedback on the stuttered voice conversion experiments in Section~\ref{subsec:practical_stutter}, and Chris-Mari Schreuder for help with the textual descriptions of voices for Section~\ref{sec:text-to-voice}.

\vspace{-3mm}
\bibliographystyle{splncs04}
\bibliography{references}

\end{document}